# Double Cross Magnetic Wall Decoupling for Quadrature Transceiver RF Array Coils using Common-Mode Differential-mode Resonators

Komlan Payne, *Graduate Student Member, IEEE,* Aditya Ashok Bhosale, *Student Member, IEEE*, and Xiaoliang Zhang, *Member, IEEE*

*Abstract*— **In contrast to linearly polarized RF coil arrays, quadrature transceiver coil arrays are capable of improving signal-to-noise ratio (SNR), spatial resolution and parallel imaging performance. Owing to a reduced excitation power, low specific absorption rate can be also obtained using quadrature RF coils. However, due to the complex nature of their structure and their electromagnetic proprieties, it is challenging to achieve sufficient electromagnetic decoupling while designing multichannel quadrature RF coil arrays, particularly at ultrahigh fields.**

**In this work, we proposed a double cross magnetic wall decoupling for quadrature transceiver RF arrays and implemented the decoupling method on common-mode differential mode quadrature (CMDM) quadrature transceiver arrays at ultrahigh field of 7T. The proposed magnetic decoupling wall comprised of two intrinsic decoupled loops is used to reduce the mutual coupling between all the multi-mode current present in the quadrature CMDM array. The decoupling network has no physical connection with the CMDMs' coils giving leverage over size adjustable RF arrays. In order to validate the feasibility of the proposed cross magnetic decoupling wall, systematic studies on the decoupling performance based on the impedance of two intrinsic loops are numerically performed. A pair of quadrature transceiver CMDMs is constructed along with the proposed decoupling network and their scattering matrix is characterized using a network analyzer. The measured results show all the current modes coupling are concurrently suppressed using the proposed cross magnetic wall. Moreover, field distribution, and SNR intensity are numerically obtained for a well-decoupled 8-channel quadrature knee-coil array.**

## I. INTRODUCTION

RADIO frequency (RF) coils are one of the most critical hardware used for in-vivo proton imaging or other NMR-sensitive nucleus applications [1-3]. The quest for more detailed images of human anatomy, has enabled the development and research of ultra-high field (UHF) MRI where the static magnetic field $B_0 \geq 7T$ [4-8]. However, many studies have demonstrated a crucial trend in nonuniformity of the field distribution obtained from RF coil due to the reduced effective wavelength of the Larmor frequency at UHF [9, 10]. To alleviate such distortion from the field distribution, RF coils array with phase and amplitude control technique were adopted to enable B1 shimming and obtain a localized homogeneous field in the targeted region [11-15]. This inquiry has gained substantial importance in MRI. The choice of the type of RF resonator, their size/shape, their position with respect to the sample, and their filling factor are all tailored for optimal signal to noise ratio (SNR) and penetration depth in the ROI [16-18]. Improved SNR and thus spatial resolution can be obtained by using a multichannel RF array which is close-fitting to the imaging sample. However, electrodynamic interaction between resonant elements of an array is a major source of degradation in the B1 field strength and the parallel imaging performance. In order to counter such problems, decoupling networks [19-24] are used to shield coil element from undesired electromagnetic existing in a high-density array. A handful of techniques have been employed in the literature to avoid electromagnetic destructive interference within an array of MRI coil. From geometry overlapping decoupling for adjacent elements [25-31], to transformers, capacitive/inductive decoupling lumped networks, passive distributed resonators between inter-elements, [19, 20, 23, 24, 32-38], metamaterial superstrate [39, 40], and high impedance coils [25], all these techniques have proved their capability to assure that RF coils in an array are well decoupled for many extremity MRI applications. In this work, we design and investigate the feasibility of a magnetic wall decoupling for quadrature transceiver RF array coils for ultrahigh field 7T MR imaging applications.

Transceiver RF coils with the capability to achieve quadrature or circularly polarized electromagnetic field have

This work was supported in part by the NIH under Grant U01 EB023829 and SUNY Empire Innovation Professorship. (Corresponding author: Xiaoliang Zhang.)

Komlan Payne is with Department of Biomedical Engineering, State University of New York at Buffalo, Buffalo, NY 14260 USA (e-mail: komlanpa@buffalo.edu).

Aditya Ashok Bhosale is with Department of Biomedical Engineering, State University of New York at Buffalo, Buffalo, NY 14260 USA (e-mail: adityaas@buffalo.edu).

Xiaoliang Zhang is with the Departments of Biomedical Engineering and Electrical Engineering, State University of New York at Buffalo, Buffalo, NY 14260 USA (e-mail: xzhang89@buffalo.edu).

demonstrated superior performance over linear polarized field, providing higher signal to noise ratio (SNR) and reduced signal excitation power [41-44]. At UHF, hardware in MRI systems has made progress toward the use of RF coils based on microstrip transmission line (MTL) theory [8] owing to their inherent low radiation losses over conventional RF coils. The well-known common mode differential mode (CMDM) resonator that incorporated two orthogonal and intrinsically decoupled modes, the common-mode (CM) and differential-mode (DM), was employed for the design of double nuclear [45] and quadrature [46] transceiver. While the CMDM resonator has served as a building block to design a volume coil for double nuclei $^1$H/$^{13}$C MR acquisitions due to the enhance coupling between elements [45], there is a unmet need for developing decoupled multichannel CMDM as quadrature transceiver arrays which in other hand will benefit SNR enhancement, image homogeneity, parallel imaging and parallel excitation applications. The implementation of quadrature transceiver arrays using CMDM resonators could be technical challenging due to the complex coupling between the two existing current modes which operate at the same resonant frequency in such resonator. In order to circumvent aforementioned electromagnetic coupling issue between a pair of closely placed CMDM resonators, a modified magnetic wall network is proposed as a non-overlapping decoupling method which can efficiently decouple this quadrature CMDM coil array. Magnetic wall decoupling technique based on induced current elimination (ICE) has previously been theoretically analyzed [47, 48] and used to diminish electromagnetic coupling between linear polarized resonator RF array for parallel magnetic resonance imaging application [19, 20, 49-52]. However, it is technically challenging to use a single magnetic wall decoupling between a pair of closely placed CMDM resonator with quadrature or circularly polarized field due to the complex nature of the coupling existing between the multi-mode current. Hence, we proposed a double cross magnetic wall decoupling technique to reduce undesired field coupling for multichannel CMDM quadrature RF coil arrays at 7 Tesla. To validate the proposed decoupling technique, an eight-channel CMDM quadrature array loaded with cylindrical phantom is designed with and without the double cross magnetic decoupling wall and their decoupling performance is numerically evaluated using full wave electromagnetic analysis (HFSS). A pair of the fabricated quadrature CMDM resonator is also tested on bench test to assess their scattering parameters with/without the decoupling wall. Furthermore, the eight-channel transceiver CMDM quadrature array with the double cross magnetic wall decoupling is loaded with a more realistic human foot phantom to evaluate inter-element decoupling, EM field maps and the local specific absorption rate (SAR) within the phantom.

## II. Description of Methodology

In this section, a single quadrature element using CMDM resonator is designed to operate at 300 MHz (the Larmor Frequency of proton 1H at 7T). The CMDM resonator based on split microstrip transmission line is made of 4 mm strip copper width (conductivity $\sigma$ = 5.87 × 10$^7$ S/m) and a ground plane separated by a dielectric substrate. The strip copper is a rectangular loop (width = 3 cm, length = 6 cm ) shorted to the ground plane by 2 stubs line. This planar configuration supports both common mode (CM) and differential mode (DM) current distributions (details can be found in [45, 46] ). A coaxial probe feeding (with the inner conductor connected directly to the center of the microstrip and the outer connected to the ground plane) is used to drive the CM port while the DM port is driving by inductive coupling using a small square loop (1.5 cm x 1.5 cm). The electric parameters of the CMDM are depicted in the caption of Fig. 1. Variable capacitors $C_1$, $C_2$, and $C_3$ are used to tune the two resonant modes to 300 MHz, while $C_{m1}$ and $C_{m2}$ are used to match both ports to 50 ohms. The CMDM resonator is supported by an acrylic cylinder (with 15 cm outer diameter and 13 cm inner diameter) loaded with a cylindrical water phantom (conductivity $\sigma$ = 0.6 S/m and permittivity $\varepsilon_r$ = 81). The diameter and length of the cylindrical water phantom are set to 13 cm and 20 cm. As shown in Fig. 1, the simulated scattering parameters of the single element CMDM resonator indicates intrinsic decoupling between both the CM and DM ports which can be used to generate quadrature field.

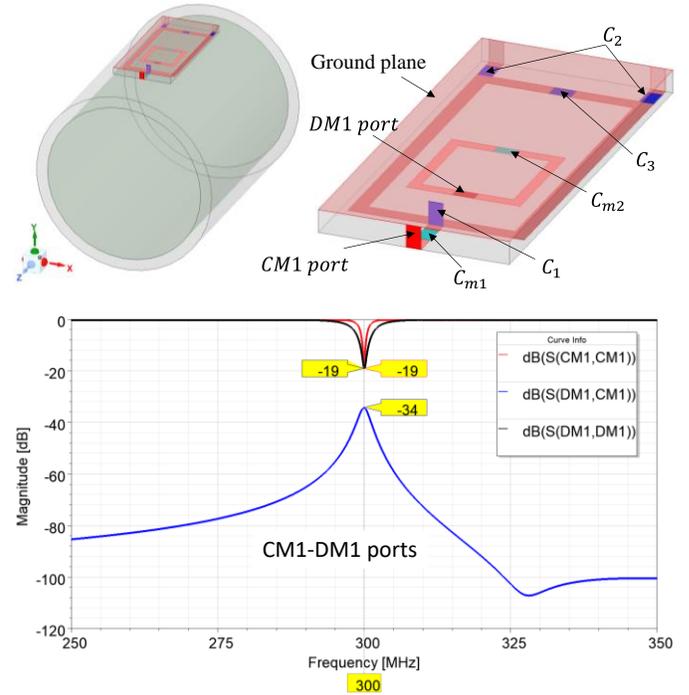

Fig. 1. Topology of the quadrature CMDM resonator with detailed electrical parameters. $C_1$= 95 pF, $C_2$ = 4.8 pF, $C_3$ = 0.5 pF, $C_{m1}$= 30 pF, $C_{m2}$= 7.7 pF. The CMDM resonator is placed 1 cm on top of a cylindrical water phantom (conductivity $\sigma$ = 0.6 S/m and permittivity $\varepsilon_r$ = 81). (b) Simulated scattering parameters of the CMDM. Good input impedance and isolation are obtained between both common mode (CM1) and differential mode (DM1) ports at the Larmor frequency of $^1$H at 7 Tesla UHF.

### A. Single magnetic wall decoupling

A pair of the CMDM resonator separated by 1.5 cm circular gap along the circumference of the acrylic cylinder is designed for parallel imaging [see Fig. 2(a)]. Due to the close-fitting array configuration, electromagnetic coupling is generated between the CM1-CM2 and DM1-DM2 ports. In order to mitigate these coupling, a single magnetic wall decoupling is placed between adjacent CMDM resonator. A magnetic wall is a resonator implemented with lumped circuit element or distributed element or a mixture of both, use to suppress the induced current between the array elements [19]. In this study we adopt a loop with tunable capacitors within the transmission line. The pair of quadrature CMDM resonator with the decoupling wall is illustrated in Fig. 2. The effectiveness of the magnetic wall to alleviate the multi-mode coupling between the resonators is investigated and its performance is compared with the pair of CMDM without the decoupling network. The scattering parameter of the pair of quadrature CMDM resonators without the magnetic wall decoupling shows that electromagnetic coupling between the two modes exists. As illustrated in Fig. 2(b), simulations results show strong coupling (with 2 split resonant frequencies) obtained between the CMs and also between the DMs ports which constitute a major factor that degrade the capability of transmission and detection of the antenna array. In order to mitigate the coupling between neighboring resonators, the ICE loop used as magnetic wall is implemented such that its impedance $Z_d$ satisfy the equation [47]:

$$Z_d = Z_{dc}^2 / Z_{cc} \quad (1)$$

where $Z_{cc}$ is the mutual impedance between CMDM1 & CMDM2 and $Z_{dc}$ is the mutual impedance between ICE & CMDM. Since the two current modes (with different impedances) exist in the single CMDM resonator, the multi-mode coupling from a pair of CMDMs can't be sufficiently suppressed concurrently with the ICE loop based on (1). Hence, we implement the magnetic wall to efficiently suppress either the common mode or the differential mode coupling to the detriment of the other mode coupling. By evaluating the impedance of the ICE loop using a tuning capacitance $C_{c1}$ with a value of 3.5 pF, the coupling between the common mode ports is suppressed (from -5 dB to -17 dB at 300 MHz) while the decoupling between the differential ports remained compromised [see Fig. 4(c)]. In the other hand the magnetic wall impedance can be tuned by setting $C_{c1}$= 0.9 pF to suppress the coupling between the differential mode ports while compromising on the common mode port decoupling [see Fig. 4(d)]. Per our analysis, the single magnetic wall decoupling is unable to concurrently alleviate the multi-mode coupling between the pair of CMDM resonators which support our previous results in [53]. The optimization of the parameters of such ICE loop, particularly its impedance, will help to further improve decoupling performance for all resonant modes.

### B. Double cross magnetic wall decoupling

Due to the inability of the single ICE loop to provide adequate decoupling of all the multi-mode coupling existing in a pair of the closely placed CMDM resonator, we introduce a double cross decoupling magnetic wall. Such decoupling network is composed of two orthogonal loops with tuning capacitors as shown in Fig 3. The two ICE loop are intrinsically

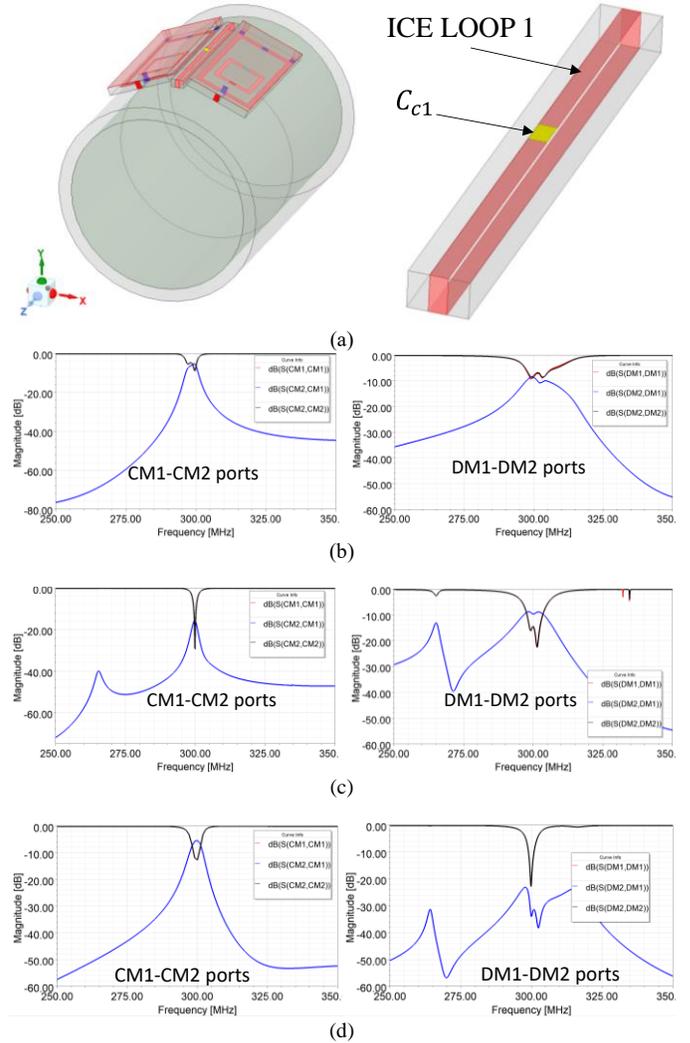

Fig. 2. (a) Design model of a pair of quadrature CMDM resonator integrated with a single magnetic loop decoupling placed 1 cm on top of the cylindrical water phantom (conductivity $\sigma$ = 0.6 S/m and permittivity $\varepsilon_r$ = 81). (b) Simulated frequency response of the CMDMs without the ICE loop. Results show strong electromagnetic coupling between the multimode resonators. (c) Simulated frequency response of the CMDMs with the ICE loop integrated ($C_{c1}$= 3.5 pF). The magnetic wall is optimized to suppress only the coupling between the common-mode ports. (d) Simulated frequency response of the CMDMs with the ICE loop integrated ($C_{c1}$= 0.9 pF). The magnetic wall is optimized to suppress only the coupling between the differential mode ports.

decoupled from each other and can be used to independently to suppress all the existing multi-mode coupling current. By doing so, ICE loop1 can be tuned to mitigate the common mode coupling while ICE loop 2 can be optimized to reduce the differential mode coupling. From (1), the impedance of the ICE loop 1 & 2 can be numerically obtained, and the value of the tuning capacitances ($C_{c1}$ and $C_{c2}$) can be further optimized to provide sufficient decoupling of all the multi-mode coupling current. As shown in Fig. 3, both common mode ports and differential mode ports are concurrently suppressed using the double cross magnetic wall with tuning capacitances $C_{c1}$ = 3.4 pF and $C_{c2}$ = 1.6 pF. The simulated scattering parameters versus frequency, show excellent matching performance at all the ports

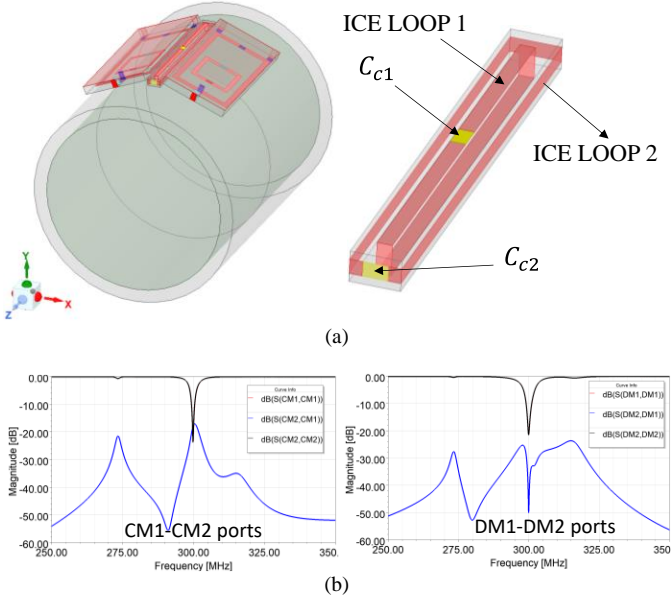

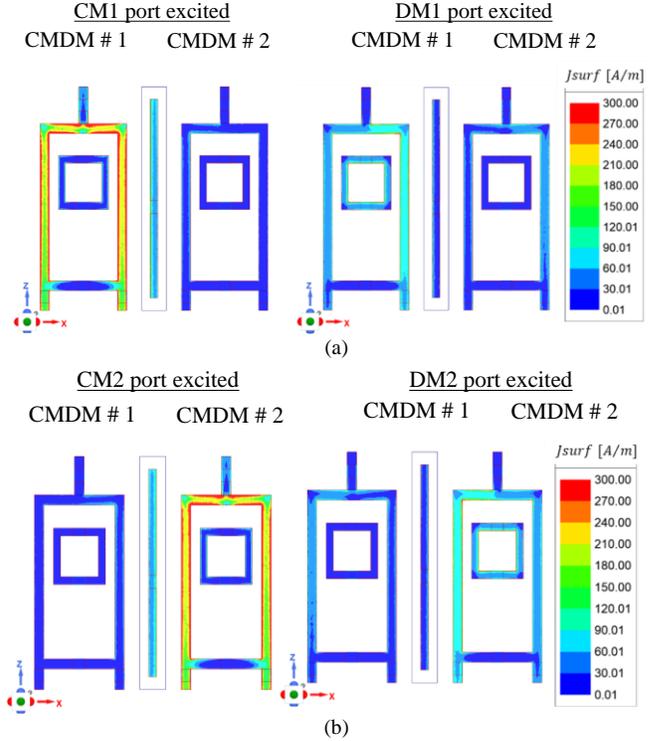

Fig. 3. (a) Design model of a pair of quadrature CMDM resonator integrated with a double cross magnetic loop decoupling ($C_{c1}$ = 3.4 pF and $C_{c2}$ = 1.6 pF) placed 1 cm on top of the cylindrical water phantom (conductivity $\sigma$ = 0.6 S/m and permittivity $\varepsilon_r$ = 81). (b) Simulated scattering parameters at the Larmor frequency showing excellent impedance match at all the ports (CMs and DMs). Good isolation is also obtained at the ports, about 19 dB and 50 dB between the CMs and DMs ports, respectively.

Fig. 4. Simulated surface current distribution obtained at 300 MHz within the conductors. (a) CMDM #1 resonator is linearly excited using CM1 port or DM1 port; (b) CMDM #2 resonator is linearly excited using CM2 port or DM2 port. In any case a very negligible residual current is obtained in the passive CMDM coil.

(< -20 dB) and strong decoupling performance. The coupling between the CM ports is suppressed about 19 dB while the coupling is reduced up to 50 dB between the DM ports leading to a very minimum power crosstalk between both CMDMs resonators.

In order to support aforementioned decoupling performance between the pair of CMDMs resonators using the double cross magnetic wall, we evaluate the current distribution within the conductors and the RF transmit $B_1^+$ field strength maps coupled to the water phantom. The simulated current density distribution of the well-decoupled pair of CMDMs at the Larmor frequency is illustrated in Fig. 4. Results are obtained when one port is excited with 1 W input-power and the other ports are terminated with a 50 ohms impedance. As can be seen in Fig. 4, two different current distribution (parallel current and loop current type) are obtained, respective to the common-mode and differential mode path. The simulated current distribution with one CMDM excited shows a very negligible residual current in the other passive CMDM coil, which is characteristic of strong decoupling performance of the double cross magnetic wall. The RF transmit field strength map ($B_1^+$) normalized to 1 W of the accepted input power of individual CMDM resonators is also obtained through simulation. Each CMDM is driving in quadrature mode using its two orthogonal ports (CM and DM) with 90-degree phase difference. As shown in Fig. 5, individual transmit $B_1^+$ field at the transversal plane of the water phantom shows a noise free distinctive image from its neighboring element, which further validate the efficacy of the double cross magnetic wall in term of high-inter RF coils isolation.

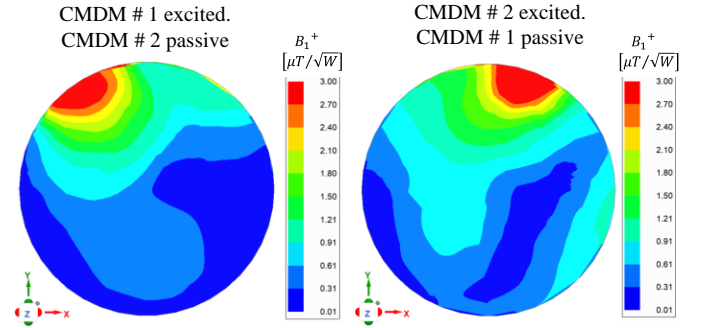

Fig. 5. Simulated transmit $B_1^+$ field distribution normalized to the accepted input power in the XY (transversal) plane through the center of the RF coils. The excited CMDM is driving with a quadrature power source (phase of CM port = 90 °, DM port = 0°)

### C. Bench test measurement of the well-decoupled CMDMs pair of resonators

A pair of the CMDM resonators is manufactured using a printed circuit board technology. The dimensions of the design board are the same used in the numerical simulations. The design pattern is printed on a low loss TLX-9 from Taconic PTFE substrates laminated with 1 oz. copper. The dielectric slabs have a relative permittivity value of $\varepsilon r$ =2.5 and a loss tangent of tanδ = 0.001. Trimmer capacitors (from Digi-Key) are soldering in the coils design for matching and tuning of the CDMDs resonator as show in Fig. 6(a). The manufactured boards are designed to operate at 7T Larmor frequency and matched to 50 ohms. The design is placed 1 cm on top of a cylindrical water phantom with roughly the same electrical parameter used in the numerical simulation. The scattering parameters of the pair of CMDM resonator with and without the decoupling mechanism is obtained using a vector network analyzer (ZVL, Rohde & Schwarz, Munich, Germany). The fabricated resonators are placed about 1 cm from each other, and the bench test results without the decoupling magnetic wall shows strong electromagnetic coupling between the CM ports as well as the DM ports [see Fig. 6 (b)]. By integrating the double cross ICE loop with optimized capacitance value ($C_{c1}$ = 3.2 pF and $C_{c2}$ = 1.4 pF), good isolation about 20 dB is achieved between all the ports as shown in Fig. 6 (c).

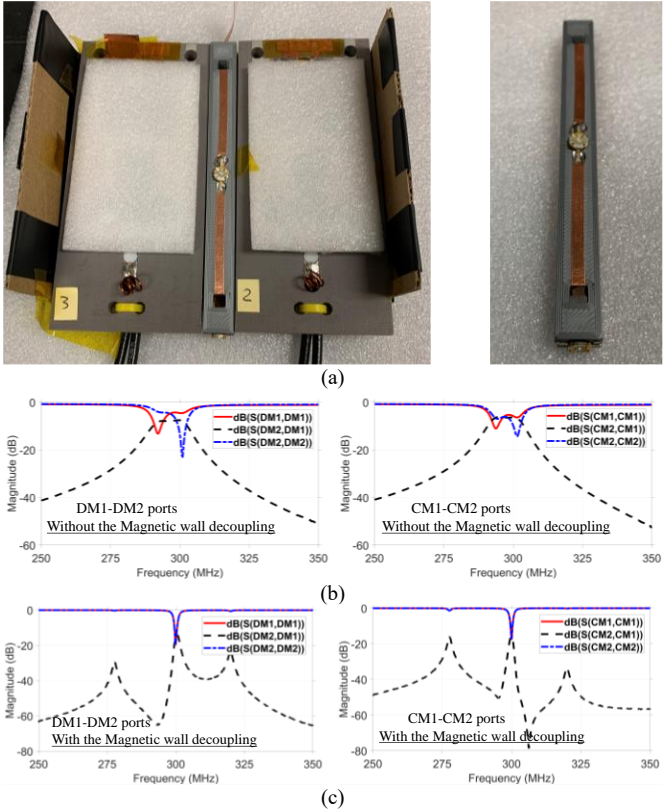

Fig. 6. (a) Photograph of the prototype pair of the fabricated CMDM coils separated by 1 cm distance from each other along with the ICE loops. (b) Measured scattering parameters of the RF coils without the magnetic wall decoupling showing strong mutual coupling between both resonators. (c) Measured scattering parameters of the RF coils with the magnetic wall decoupling showing excellent matching (about -20 dB) and good isolation (about 18 dB) are achieved between DM1-DM2 ports as well as the CM1-CM2 ports.

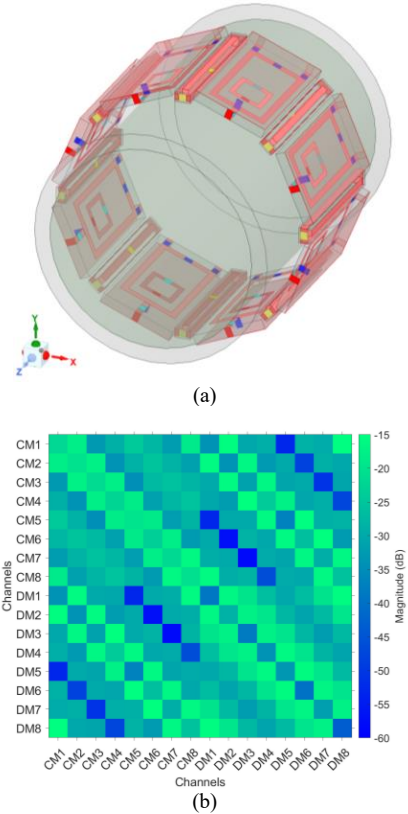

Fig. 7. (a) Configuration of the close-fitting 8-channel CMDM array integrated with the proposed cross magnetic wall decoupling. (b) Scattering matrix of the array showing a well-decoupled resonators and good matching performance of the design.

### III. WELL-DECOUPLED 8 CHANNELS QUADRATURE CMDM ARRAY FOR FIELD AND IMAGE ANALYSIS

For MRI development, we evaluate the performance of an array of 8 channels quadrature CMDM resonators. The double cross magnetic wall is placed between adjacent elements for decoupling purpose. The array is wrapped on an acrylic cylinder (with 15 cm outer diameter and 13 cm inner diameter) that enclosed a cylindrical water phantom (conductivity $\sigma$ = 0.6 S/m and permittivity $\varepsilon_r$ = 81). The circular gap along the circumference of the acrylic cylinder between adjacent CMDM resonators is about 1.5 cm. The elements in the array configuration illustrated in Fig. 7 are matched to 50 ohms and tuned to operate at 300 MHz. The scattering matrix of the 8-channels is depicted in Tables I. The simulated results show good impedance matching ( < -18 dB) for all the channels and good isolation between all the ports (< -15 dB). Each resonator is excited with a quadrature source and individual field map of each channel at the transverse plane is computed. Each field distribution is obtained for the active channel while other channels are terminated with 50 ohms impedances. The individual transmits $B_1^+$ field distribution normalized to 1 W of the accepted input power of each channel shows a distinctive image at the center axial plane of the water phantom as seen in Fig. 8. These field maps with no distortion reflect the strong decoupling between neighboring CMDM resonators. For the combined transmit field, all the 8 quadrature channels are simultaneously excited with a 45° linear phase progression (CP B1 shimming mode) between neighboring CMDM resonators.

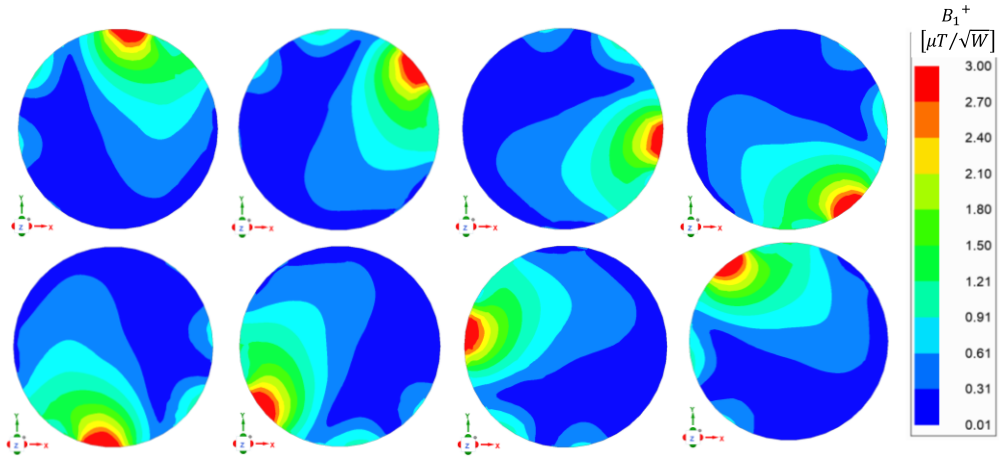

Fig. 8. Individuals transmit ($B_1^+$) field distribution of each channel excited with 1 W input quadrature power source on the water phantom at the axial plane through the center of the RF coils . Each field distribution is obtained for the active channel while other channels are terminated with 50 ohms impedances.

A constructive transmit field $B_1^+(total)$ is obtained at the center of the phantom [see Fig. 9(a)]. Strong intensity field is also obtained at the peripheral of the water phantom. For safety requirement due human exposure from high power electromagnetic pulse [54], the local specific absorption rate (SAR) averaged over 10 g is also numerically computed for the well-decoupled/matched 8 channels quadrature CMDM resonators. Simulated results of the water phantom in Fig. 9(b) show that the peak SAR value at central axial slice is about 2 W/Kg and well below the FDA limit (10 W/Kg).

By leaning toward a more realistic analysis, the proposed quadrature RF array coils with integrated double cross magnetic wall decoupling is used to produce detailed images of a knee from a build-in human right foot voxel model from ANSYS HFSS. The configuration of the foot voxel with the quadrature array coils is illustrated in Fig. 10(a). The average diameter of the foot voxel model is about 13 cm. However, due to the asymmetric shape of the foot, the distance between the coils and the knee varies from 1 to 1.4 cm. Due to the load sensitivity, a fine-tuning of the parameters is needed to optimize the design performance. The electrical parameters of the design are depicted in Fig. 10 caption. The elements in the array resonator are excited using 50 ohms lumped ports with 1 W quadrature power source and 45° linear phase progression (circular polarized shimming mode). The scattering parameter matrix of the 8 channels illustrated in Fig. 10(b) shows show good matching performance (< -15 dB for all the channels) and strong decoupling performance between all the channels in the array which greatly benefit parallel imaging and parallel excitation. The computed total RF transmit/receive ($B_1^+/B_1^-$) efficiency show high intensity field in the ROI (knee imaged) of the foot voxel as can be seen in Fig. 10(c) & (d). For safety and reliability considerations, both 3D and 2D local SAR distribution are also simulated and illustrated in Fig. 10(e) & (f). Few hot spots can be seen around the knee due to enhanced localized electric field intensity in the tissue losses. Yet, the peak SAR value is considerably low about 2.3 W/Kg.

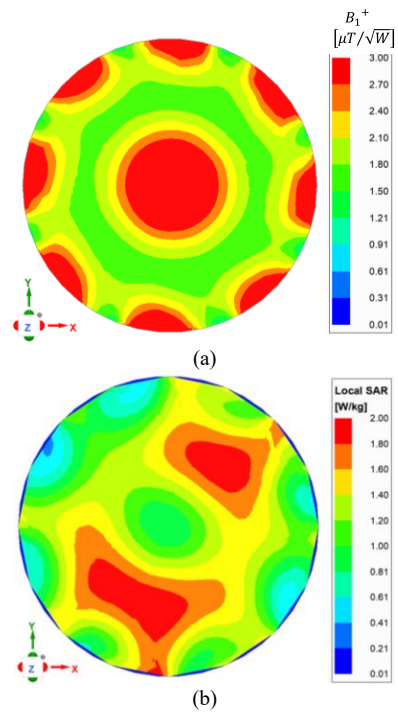

Fig. 9. All the 8 channels quadrature CMDM resonators are simultaneously excited with 45° linear phase progression. (a) Simulated combined transmit $B_1^+(total)$ and local $SAR_{10g}$ field distribution on the water phantom at the transversal plane through the center of the RF coils. (b) Simulated SAR map within the water phantom obtained at central slice.

## IV. CONCLUSION

We have proposed a double cross magnetic wall decoupling for quadrature transceivers RF array coils based on common mode differential resonators for parallel imaging application. The proposed decoupling network composed of two orthogonal loops with tuning capacitors is analyzed to concurrently suppress all the existing multi-mode current coupling in a close-fitting array of 8 quadrature resonators. Bench tests results have demonstrated the feasibility of this decoupling mechanism for the quadrature CMDM resonator. This efficient decoupling technique is expected to be provide a more flexible design control of the undesirable coupling for non-overlapped quadrature transceivers array (such as microstrip + loop, dipole + loop, or microstrip loop + dipole types). Possibly, by evaluating the impedance of such doubled cross magnetic wall decoupling, it can also be used to mitigate inter-coupling element for double-tuned transceiver array.

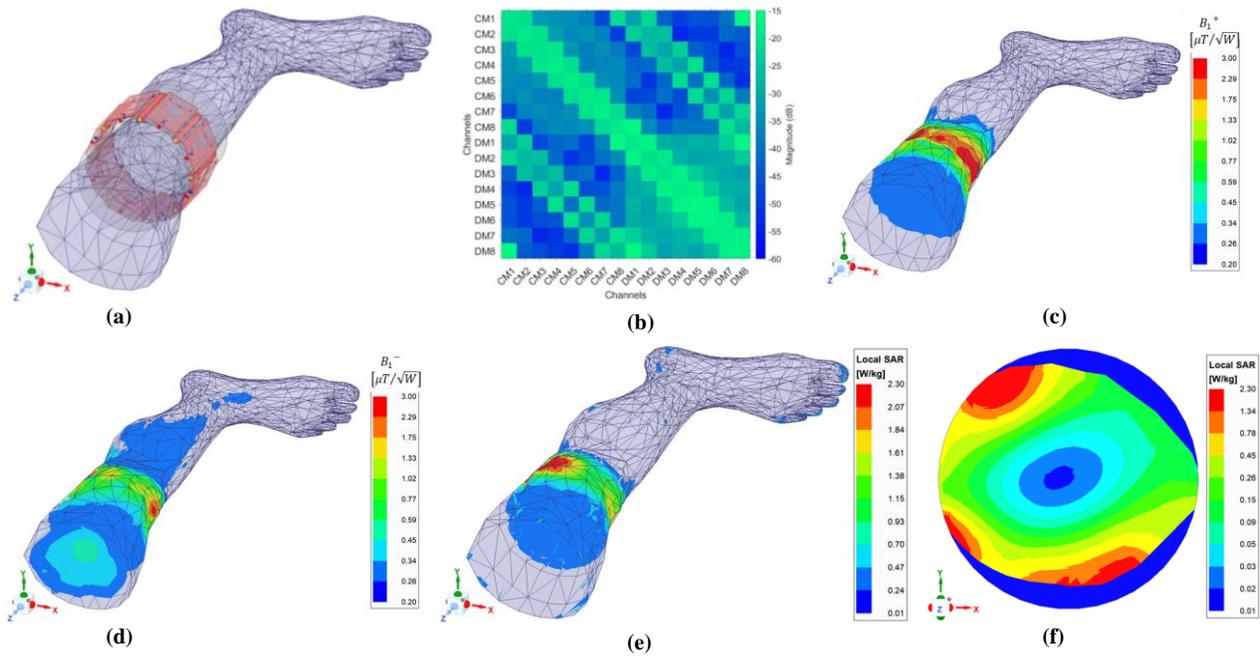

Fig. 10. (a) Design model of the quadrature 8 CMDMs array resonators used for human knee imaged. The electrical parameters for optimized performance: $C_1$= 115 pF, $C_2$ = 4.8 pF, $C_3$ = 0.3 pF, $C_{m1}$= 4 pF, $C_{m2}$= 8 pF, $C_{c1}$ = 3.4 pF, and $C_{c2}$ = 0.8 pF. (b) Simulated scattering parameters of the array showing good input impedance matching and excellent decoupling among all the ports at 300 MHz. (c-d) 3D map of the simulated transmit/receive ($B_1^+/B_1^-$) RF field distribution. (e-f) 3D & 2D map of the simulated local $SAR_{10g}$ on the foot voxel model obtained from the 8-element arrays excited with 1 W input quadrature power source and 45° linear phase progression.

ACKNOWLEDGMENT

This work is supported in part by the NIH under a BRP grant U01 EB023829 and SUNY Empire Innovation Professorship Award.